\documentclass[aps,prd,floatfix,12pt]{revtex4}
\usepackage{epsfig}
\usepackage{bm}
\usepackage{dcolumn}
\usepackage{latexsym}

\begin{document}
   
\preprint{\rightline{ANL-HEP-PR-10-45}}
   
\title{Thermodynamics of lattice QCD with 2 sextet quarks on $\bm{N_t=8}$ 
lattices}

\author{J.~B.~Kogut}
\affiliation{Department of Energy, Division of High Energy Physics, Washington,
DC 20585, USA}
 \author{\vspace{-0.2in}{\it and}}
\affiliation{Dept. of Physics -- TQHN, Univ. of Maryland, 82 Regents Dr.,
College Park, MD 20742, USA}
\author{D.~K.~Sinclair}
\affiliation{HEP Division, Argonne National Laboratory, 9700 South Cass Avenue,
Argonne, IL 60439, USA}

\begin{abstract}
We continue our lattice simulations of QCD with 2 flavours of colour-sextet
quarks as a model for conformal or walking technicolor. A 2-loop perturbative
calculation of the $\beta$-function which describes the evolution of this 
theory's running coupling constant predicts that it has a second zero at a
finite coupling. This non-trivial zero would be an infrared stable fixed point,
in which case the theory with massless quarks would be a conformal field
theory. However, if the interaction between quarks and antiquarks becomes
strong enough that a chiral condensate forms before this IR fixed point is
reached, the theory is QCD-like with spontaneously broken chiral symmetry and
confinement. However, the presence of the nearby IR fixed point means that 
there is a range of couplings for which the running coupling evolves very
slowly, i.e. it 'walks'. We are simulating the lattice version of this theory
with staggered quarks at finite temperature studying the changes in couplings
at the deconfinement and chiral-symmetry restoring transitions as the temporal
extent ($N_t$) of the lattice, measured in lattice units, is increased. Our 
earlier results on lattices with $N_t=4,6$ show both transitions move to
weaker couplings as $N_t$ increases consistent with walking behaviour. In this
paper we extend these calculations to $N_t=8$. Although both transition again
move to weaker couplings the change in the coupling at the chiral transition
from $N_t=6$ to $N_t=8$ is appreciably smaller than that from $N_t=4$ to
$N_t=6$. This indicates that at $N_t=4,6$ we are seeing strong coupling effects
and that we will need results from $N_t > 8$ to determine if the
chiral-transition coupling approaches zero as $N_t \rightarrow \infty$, as
needed for the theory to walk.
\end{abstract}

\maketitle

\section{Introduction}

We are interested in extensions of the standard model which have a 
strongly-coupled (composite) Higgs sector. The most promising theories of
this type are the so-called technicolor theories
\cite{Weinberg:1979bn,Susskind:1978ms}, QCD-like gauge theories
with massless (techni-)quarks, where the (techni-)pions play the role of the 
Higgs field, giving masses to the $W$ and $Z$. Such theories tend to have
phenomenological problems, especially when they are extended to give masses to
quarks and leptons. Walking technicolor theories, where gauge group and fermion 
content are chosen so that the running coupling constant evolves very slowly 
(`walks'), might be able to avoid such difficulties
\cite{Holdom:1981rm,Yamawaki:1985zg,Akiba:1985rr,Appelquist:1986an}. 
Deciding whether a candidate gauge theory has the properties needed is a
non-perturbative question. Hence lattice gauge theory simulation methods are
the only way to answer this reliably.

For a given gauge group with $N_f$ fermions in a specified representation of
that group, there is some value of $N_f$, below which the gauge theory is 
asymptotically free. Below this value there is a range of $N_f$ for which the
second term in the perturbative Callan-Symanzik $\beta$-function has the
opposite sign from the first. Hence, if the 2-loop $\beta$ function describes
the physics the theory with $N_f$ in this range, $\beta$ has a second
non-trivial zero representing an infrared (IR) fixed point. If this is true
the theory is a conformal field theory with a continuous spectrum. However,
there is a second possibility. If the fermion-antifermion coupling becomes
strong enough that a chiral condensate forms before the would-be fixed point
is reached, this effectively removes the fermions from consideration for
longer distances, the IR zero is avoided, and the coupling approaches infinity
at large distances. In this case the theory is QCD-like with confinement as
well as chiral-symmetry breaking. However, the presence of the nearby IR fixed
point means that the $\beta$ function becomes small at some value of the
coupling, and the coupling constant `walks'.

If we restrict ourselves to $SU(N_c)$ gauge groups with $N_c$ relatively small,
there are a limited number of potential candidates. These have been identified
and rough estimates of the value of $N_f$, which separates conformal from
walking behaviour, have been made \cite{Dietrich:2006cm,Appelquist:1988yc,
Sannino:2004qp,Poppitz:2009uq,Armoni:2009jn,Ryttov:2007cx,Antipin:2009wr,
Mojaza:2010cm}. 
Extensive lattice studies have been made for $N_c=3$ with fermions in the
fundamental representation of the colour group.
\cite{Kogut:1985pp,Fukugita:1987mb,Ohta:1991zi,Kim:1992pk,Brown:1992fz,
Iwasaki:1991mr,Iwasaki:2003de,Damgaard:1997ut,Deuzeman:2008sc,Deuzeman:2009mh,
Deuzeman:2010fn,Appelquist:2009ty,Appelquist:2007hu,Jin:2008rc,Jin:2009mc,
Jin:2010vm,Fodor:2009wk,Fodor:2009ff,Fodor:2011tu,Yamada:2009nt,
Appelquist:2010xv,Hasenfratz:2010fi}
There have also been studies with $N_c=2$ and fermions in the fundamental
representation of colour \cite{Kogut:1985pp,Bursa:2010xn,Ohki:2010sr}
as well as studies with $N_c=2$ and fermions in the adjoint
(symmetric tensor) representation 
\cite{Catterall:2007yx,Catterall:2008qk,Catterall:2009sb,DelDebbio:2008zf,
DelDebbio:2009fd,Bursa:2009we,Hietanen:2008mr,Hietanen:2009az,Matsufuru:2010zz,
DeGrand:2011qd}. 
Finally, there have been studies with $N_c=3$ and fermions in the sextet
(symmetric-tensor) representation of the gauge group.
\cite{Shamir:2008pb,DeGrand:2008kx,DeGrand:2009hu,DeGrand:2010na,Kogut:2010cz,
Sinclair:2010be,Fodor:2008hm,Fodor:2011tw}.

We are concentrating our efforts on QCD ($N_c=3$) with colour-sextet quarks.
For this choice, asymptotic freedom is lost at $N_f=3\frac{3}{10}$. This means
that only $N_f=2,3$ are of interest. Both of these have $\beta$-functions where
the 1 and 2-loop terms are of opposite sign. $N_f=3$ is close enough to the
number of flavours for which asymptotic freedom is lost that the IR fixed
point occurs for very weak coupling, for which perturbation theory can probably
be trusted. Hence it is believed that this theory is almost certainly a 
conformal field theory. Estimates of the value of $N_f$, which separates 
conformal from walking behaviour, suggest that $N_f=2$ is a good candidate for
walking behaviour. However, there is enough uncertainty in these methods that
a more reliable study of the $N_f=2$ is waranted. We have thus chosen to study
the $N_f=2$ theory, using lattice gauge simulations with staggered quarks.
Lattice studies of QCD with 2 flavours of colour-sextet Wilson quarks have been
performed by Degrand, Shamir and Svetitsky. To date these have been unable to
tell unambiguously whether this theory is conformal or walking. The Lattice
Higgs Collaboration have been studying this theory using improved staggered
quarks. Recently they have reported evidence that this theory spontaneously
breaks chiral symmetry which would indicate that it has walking behaviour
(unless there is a bulk chiral transition at even weaker coupling). 
%As of Lattice 2010, they have also been unable to determine the nature
%of this 2-flavour theory.

Whereas the other groups have concentrated their efforts on determining the
nature of QCD with 2 sextet quarks from studies of the zero temperature 
behaviour of the theory (apart from some early thermodynamics simulations
by Degrand, Shamir and Svetitsky), we are studying the thermodynamics of
this theory. Here we are measuring the dependence of the lattice (bare) coupling
at the deconfinement and chiral-symmetry restoration transitions on $N_t$, the 
temporal extent of the lattice in lattice units. If these are indeed
finite-temperature transitions, the couplings at which they occur should tend
to zero as $N_t \rightarrow \infty$ in a manner controlled by asymptotic
freedom. Such behaviour would indicate walking. If the theory is conformal,
these couplings should approach a non-zero constant as $N_t \rightarrow \infty$,
indicating a bulk transition. Simulations at $N_t=4$ and $6$, reported in our
earlier publication showed that both transition couplings did decrease with
increasing $N_t$. This work reports the results of simulations at $N_t=8$.
While both transitions do tend to weaker couplings as $N_t$ goes from $6$ to
$8$, the change in coupling at the chiral transition, which occurs at a 
considerably weaker coupling that the deconfinement transition, is much smaller
than that between $N_t=4$ and $6$. (Such separation of the deconfinement and
chiral-symmetry restoration transitions, which is not observed for fundamental
quarks, has been observed with adjoint quarks 
\cite{Karsch:1998qj,Engels:2005te}).
The most likely interpretation is that between $N_t=4$ and $6$, this
transition is in the strong-coupling domain where the quarks have condensed to
form a chiral condensate at length scales of order of the lattice spacing and
do not participate in the running of the coupling constant, which now runs as
in quenched QCD. Between $N_t=6$ and $8$ the chiral transition coupling
finally emerges into the weak-coupling regime where the quarks also particpate
in the running of the coupling constant. This means that we will need to
simulate at even larger $N_t$s to determine whether this theory is conformal
or walking.

As for $N_t=6$, the $N_t=8$ lattice shows a clear 3-state signal above the 
deconfinement transition. These states are characterized by the phase of the
Wilson Line (Polyakov Loop) having the values $0$, $\pm \frac{2\pi}{3}$, a
vestige of the $Z_3$ colour symmetry of the quenched theory. Within the
limitations of our simulations, all 3 states appear stable. At even weaker
couplings -- close to the chiral transition -- the 2 states with complex
phases disorder to a phase with a negative Wilson Line. This phase structure,
which is richer than that for fundamental quarks, where the Wilson Line is
always real and positive, was predicted by Machtey and Svetitsky and observed
in their simulations with Wilson quarks \cite{Machtey:2009wu}.

In section~2 we describe our simulation techniques, and how one can measure the
running of the coupling constant from thermodynamics. Section~3 describes our
simulations and results. Finally in section~4 we discuss our results, draw
conclusions, and indicate directions for future investigations.

\section{Methodology}

For the gauge fields we use the standard Wilson (plaquette) action:
\begin{equation}
S_g=\beta \sum_\Box \left[1-\frac{1}{3}{\rm Re}({\rm Tr}UUUU)\right].
\end{equation}
For the fermions we use the unimproved staggered-quark action:
\begin{equation}
S_f=\sum_{sites}\left[\sum_{f=1}^{N_f/4}\psi_f^\dagger[D\!\!\!\!/+m]\psi_f
\right],
\end{equation}
where $D\!\!\!\!/ = \sum_\mu \eta_\mu D_\mu$ with 
\begin{equation}
D_\mu \psi(x) = \frac{1}{2}[U^{(6)}_\mu(x)\psi(x+\hat{\mu})-
                            U^{(6)\dagger}_\mu(x-\hat{\mu})\psi(x-\hat{\mu})],
\end{equation}
where $U^{(6)}$ is the sextet representation of $U$, i.e. the symmetric part of
the tensor product $U \otimes U$. When $N_f$ is not a multiple of $4$ we 
use the fermion action:
\begin{equation}
S_f=\sum_{sites}\chi^\dagger\{[D\!\!\!\!/+m][-D\!\!\!\!/+m]\}^{N_f/8}\chi.
\end{equation}
The operator which is raised to a fractional power is positive definite and
we choose the real positive root. This yields a well-defined operator. We 
assume that this defines a sensible field theory in the zero lattice-spacing
limit, ignoring the rooting controversy. (See for example \cite{Sharpe:2006re}
for a review and guide to the literature on rooting.) 

We use the RHMC method for our simulations \cite{Clark:2006wp}, where the
required powers of the quadratic Dirac operator are replaced by diagonal
rational approximations, to the desired precision. By applying a global
Metropolis accept/reject step at the end of each trajectory, errors due to the
discretization of molecular-dynamics time are removed.

Finite temperature simulations are performed by using a lattice of finite
extent $N_t$ in lattice units in the Euclidean time direction, and of infinite
extent $N_s$ in the spatial direction. In practice this means we choose
$N_s \gg N_t$. The temperature $T=1/N_ta$, where $a$ is the lattice spacing.
(In our earlier equations we set $a=1$.) Since the deconfinement temperature 
$T_d$ and the chiral symmetry restoration temperature $T_\chi$ should not
depend on $a$, and since $a=1/N_tT$, measuring the coupling $g$ at $T_d$ or
$T_\chi$ as a function of $N_t$ gives $g(a)$ for a series of $a$ values which
approach zero as $N_t \rightarrow \infty$. If the ultraviolet behaviour of the
theory is governed by asymptotic freedom, $g(a)$ should approach zero as 
$a \rightarrow 0$, i.e. $N_t \rightarrow \infty$. The way $g_d$ and $g_\chi$
approach zero should be determined by the perturbative $\beta$ function. The
2-loop $\beta$ function
\begin{equation}
\beta(g) = -b_1 g^3 - b_2 g^5.
\end{equation}
Then expressing our coupling constant evolution in terms of $\beta=6/g^2$ (We
apologize for the fact that we are using $\beta$ for 2 different purposes)
\begin{equation}
\Delta\beta(\beta) = \beta(a) - \beta(\lambda a)
                   = (12b_1 + 72b_2/\beta)\ln(\lambda)
\label{eqn:deltabeta}
\end{equation}
through this order. For $N_f$ flavours of sextet quarks,
\begin{eqnarray}
b_1 &=& \left(11 - \frac{10}{3}N_f\right)/16\pi^2 \nonumber \\
b_2 &=& \left(102 - \frac{250}{3}N_f\right)/(16\pi^2)^2
\end{eqnarray}

If, on the other hand, the $N_f=2$ theory is conformal, the continuum, zero
coupling ($\beta \rightarrow \infty$) limit has an unbroken chiral symmetry
(and is unconfined). Hence there will be a bulk chiral transition at a finite
coupling, which survives in the $N_t \rightarrow \infty$ limit, so the coupling
and hence $\beta$ at the chiral transition will tend to a finite value in this
limit. (Since the $\beta$ value at the deconfinement transition ($\beta_d$)
is expected to be less than that at the chiral transition ($\beta_\chi$), it
follows that $\beta_d$ will also approach a finite value as 
$N_f \rightarrow \infty$.)

We determine the position of the deconfinement transition as that value of
$\beta$ where the magnitude of the triplet Wilson Line (Polyakov Loop) increases
rapidly from a very small value as $\beta$ increases. The chiral phase 
transition is at that value of $\beta$ beyond which the chiral condensate
$\langle\bar{\psi}\psi\rangle$ vanishes in the chiral limit. Because we are
forced to simulate at finite quark mass, this value is difficult to determine
directly. We therefore estimate the position of the chiral transition by
determining the position of the peak in the chiral susceptibility 
$\chi_{\bar{\psi}\psi}$ as a function of quark mass, and extrapolating to
zero quark mass. The chiral susceptibility is given by 
\begin{equation}
\chi_{\bar{\psi}\psi} = V\left[\langle(\bar{\psi}\psi)^2\rangle
                      -        \langle\bar{\psi}\psi\rangle^2\right]
\label{eqn:chi}
\end{equation}
where the $\langle\rangle$ indicates an average over the ensemble of gauge
configurations and $V$ is the space-time volume of the lattice. Since the
fermion functional integrals have already been performed at this stage, this
quantity is actually the disconnected part of the chiral susceptibility. Since
we use stochastic estimators for $\bar{\psi}\psi$, we obtain an unbiased
estimator for this quantity by using several independent estimates for each
configuration (5, in fact). Our estimate of $(\bar{\psi}\psi)^2$ is then given
by the average of the (10) estimates which are `off diagonal' in the noise.

Our $N_t=8$ simulations are performed on $16^3 \times 8$ lattices. Near the
chiral transition, where finite size effects are a concern, we also perform
simulations on a $24^3 \times 8$ lattice for the lowest quark mass. We perform
simulations with quark masses $m=0.005$, $m=0.01$ and $m=0.02$ in lattice
units, to enable continuation to the chiral ($m=0$) limit. (Since we do not have
any zero temperature measurements, the more desirable method of choosing lines 
of constant physics is impossible.) Our trajectory length is chosen to be
$\Delta\tau = 1$ where $\tau$ is the molecular-dynamics `time' in HEMCGC
normalization \cite{Bitar:1990cb}.

A more detailed discussion of our methods of choosing parameters, run lengths, 
etc. is given in our earlier paper describing our $N_t=4,6$ simulations
\cite{Kogut:2010cz}.

\section{Simulations and Results}

We simulate QCD with 2-flavours of colour-sextet staggered quarks on 
$16^3 \times 8$ and $24^3 \times 8$ lattices. For the smaller lattice we
perform simulations with masses $m=0.005$, $m=0.01$ and $m=0.02$ to allow
extrapolation to the chiral limit, for a set of $\beta$ values covering the
range $5.5 \le \beta \le 7.4$. To probe the various phases of the Wilson Line,
we use 2 different sets of runs. In the first set of runs we use an ordered
start, in which the gauge fields are set to the unit matrix on all links, at
the highest $\beta$, and use configurations from higher $\beta$s to start runs
at lower $\beta$s. The second set of runs uses a start in which the gauge 
fields are set to the unit matrix, except for the timelike gauge fields on a
single timeslice, which are set to the matrix ${\rm diag}(1,-1,-1)$. This
puts the system in a state with a real negative Wilson Loop at large $\beta$s.

The length of a typical run at a fixed $(\beta,m)$ away from the transitions is
10,000 trajectories. Close to the deconfinement transition, this is increased
to 50,000 trajectories. Run lengths of 50,000 trajectories are also used close
to the transition from a state where the Wilson Line has phase $\pm 2\pi/3$ to
one where it has phase $\pi$. We have detailed our run lengths in the appendix.

Since finite (spatial) volume effects are most likely to be present in the
weak-coupling domain at small quark masses, where they have the potential to
shift the chiral transition, we have also performed a set of simulations on 
$24^3 \times 8$ lattices at the lowest quark mass. These simulations at
$m=0.005$ cover the range $6.2 \le \beta \le 7.4$ with mesh $\delta\beta=0.1$,
and with 10,000 trajectories at each $\beta$, from positive Wilson line
starts.

\subsection{Results}

Starting from large $\beta$ values, the runs which start from a completely
ordered state with Wilson Line $+3$ continue to have positive Wilson Lines
down to $\beta=5.8$ for $m=0.02$, and down to $\beta=5.7$ for $m=0.01$ and
$m=0.005$. Below these $\beta$ values, which are just above the deconfinement
transition, we see a clear 3-state signal, where the system tunnels between
states where the Wilson Line has phases $0$, $\pm 2\pi/3$. Because of this, we
bin our data according to the phase $\phi$ of the Wilson Line for each 
configuration. Configurations where $-\pi/3 < \phi < \pi/3$ are considered to
be in the $\phi=0$ bin. Outside of this range the configurations are considered
to be in the $\pm 2\pi/3$ bins depending on whether the imaginary part of the
Polyakov loop is positive or negative. These last 2 bins are combined by
complex conjugating those Wilson Lines which have negative imaginary parts.

Starting from large $\beta$s, in those runs which start from the second 
ordering with Wilson Line $-1$, the Wilson Line remains negative down to 
$\beta \approx 6.9$ for $m=0.02$, $\beta \approx 6.8$, $m=0.01$ and 
$\beta \approx 6.7$, $m=0.005$. Below these values the system makes a
transition to a state with Wilson Line phase $\pm 2\pi/3$. Below these $\beta$
values, these runs remain in states with Wilson Line phases $\pm 2\pi/3$ down
to $\beta=5.8$, for each $m$. For $\beta=5.7$ and below we see clear 3-state
signals where the system tunnels between the 3 states. For this reason we
again bin our data according to the Wilson Line phase, $\phi$.

\begin{figure}[htb]
\epsfxsize=4.5in
\epsffile{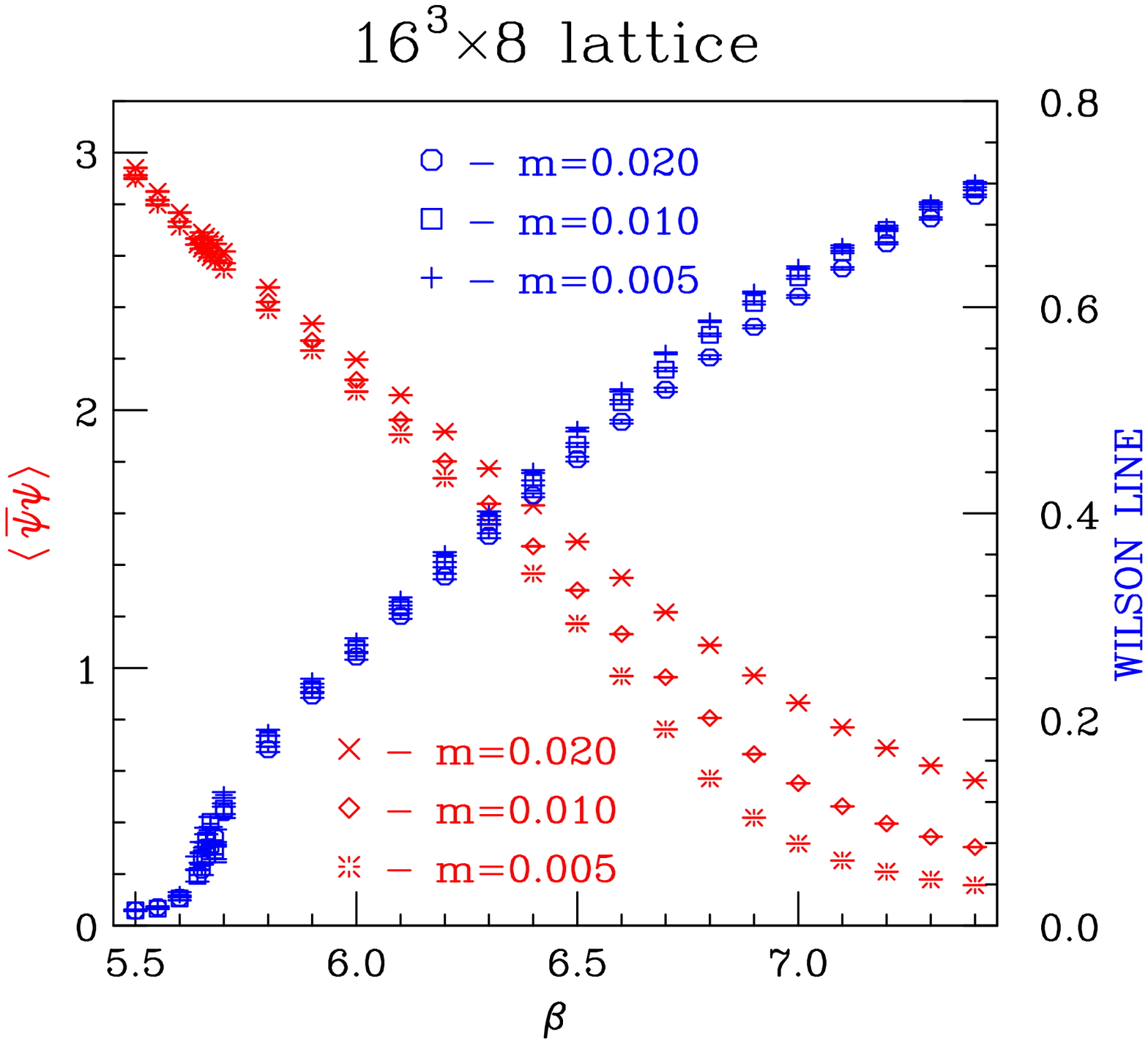}
\epsfxsize=4.5in
\epsffile{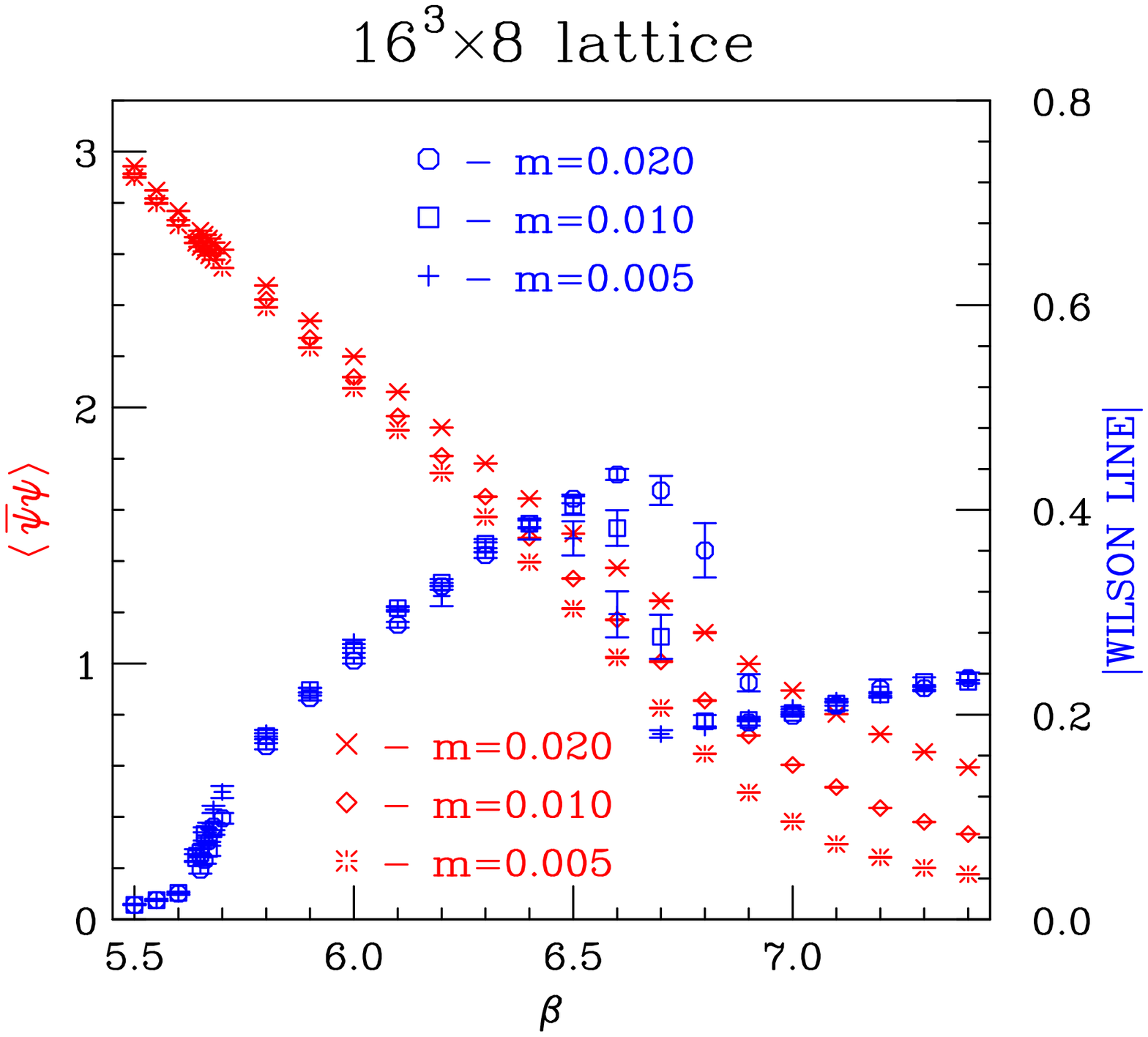}
\caption{a) Wilson Line and chiral condensate for real positive Wilson Line
states as functions of $\beta$ on a $16^3 \times 8$ lattice. \newline
b) Magnitude of Wilson Line and chiral condensate for states with complex
or real negative Wilson Lines as functions of $\beta$ on a $16^3 \times 8$ 
lattice.}
\label{fig:wil-psi}
\end{figure}

In figure~\ref{fig:wil-psi}a we present the Wilson Line and chiral condensate
for the states with a real positive Wilson Line plotted against $\beta$, for
each of the 3 masses. In figure~\ref{fig:wil-psi}b we plot the magnitude of the
Wilson Line and the chiral condensate for those states with complex or real
negative Wilson Lines.

In both sets of graphs in figure~\ref{fig:wil-psi} we observe a rapid increase 
in the (magnitude of the) Wilson Line at $\beta \approx 5.65$, corresponding to
the deconfinement transition. The sudden drop in the magnitudes of the Wilson
Lines of figure~\ref{fig:wil-psi}b for $6.6 \lesssim \beta \lesssim 7.0$ marks 
the transition where states with complex Wilson Lines ($\phi = \pm 2\pi/3$) 
disorder to a state with a real negative Wilson Line $\phi=\pi$.

\begin{figure}[htb]
\vspace*{-0.2in}
\epsfxsize=4.2in
\epsffile{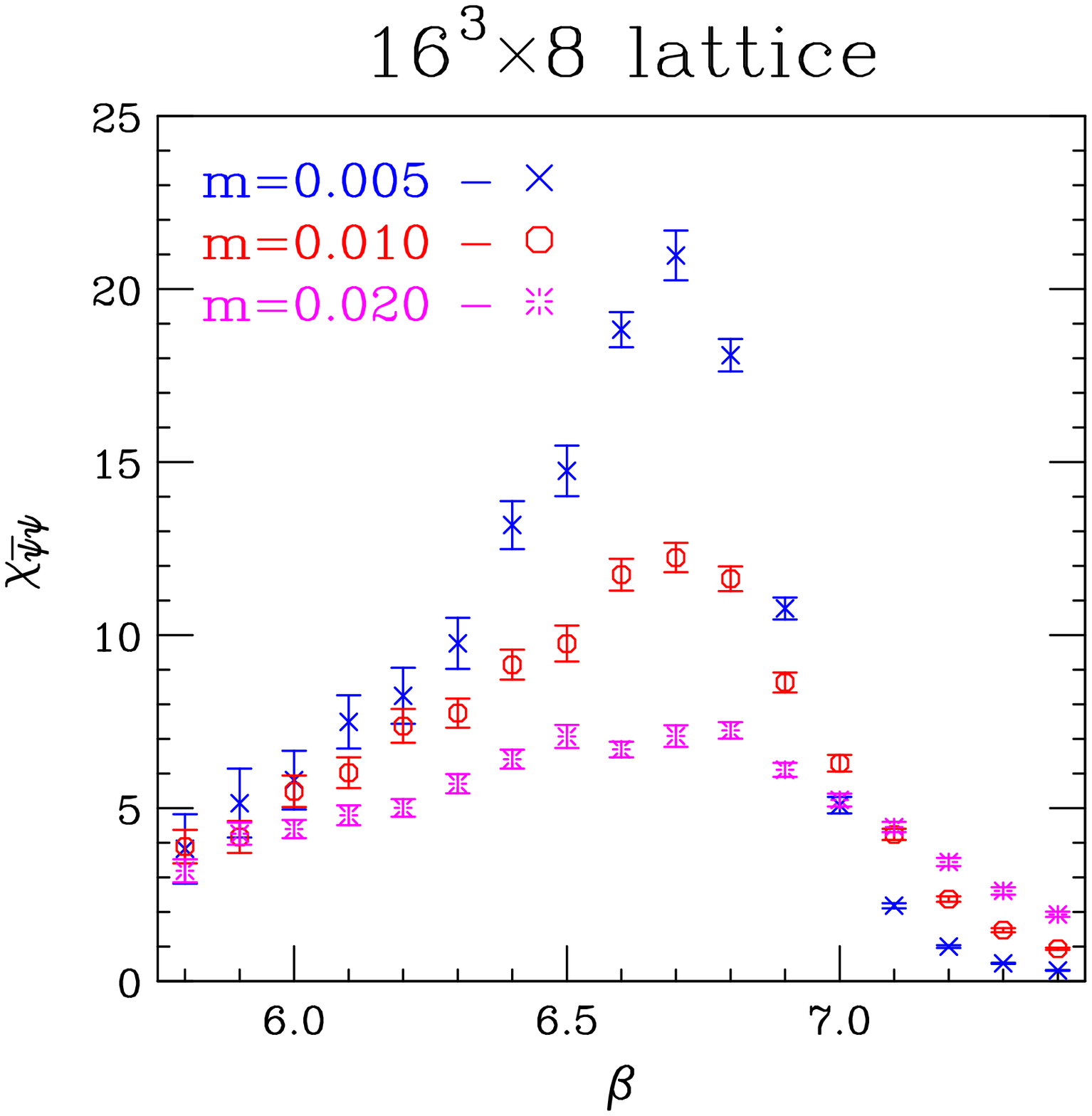}
\epsfxsize=4.2in
\epsffile{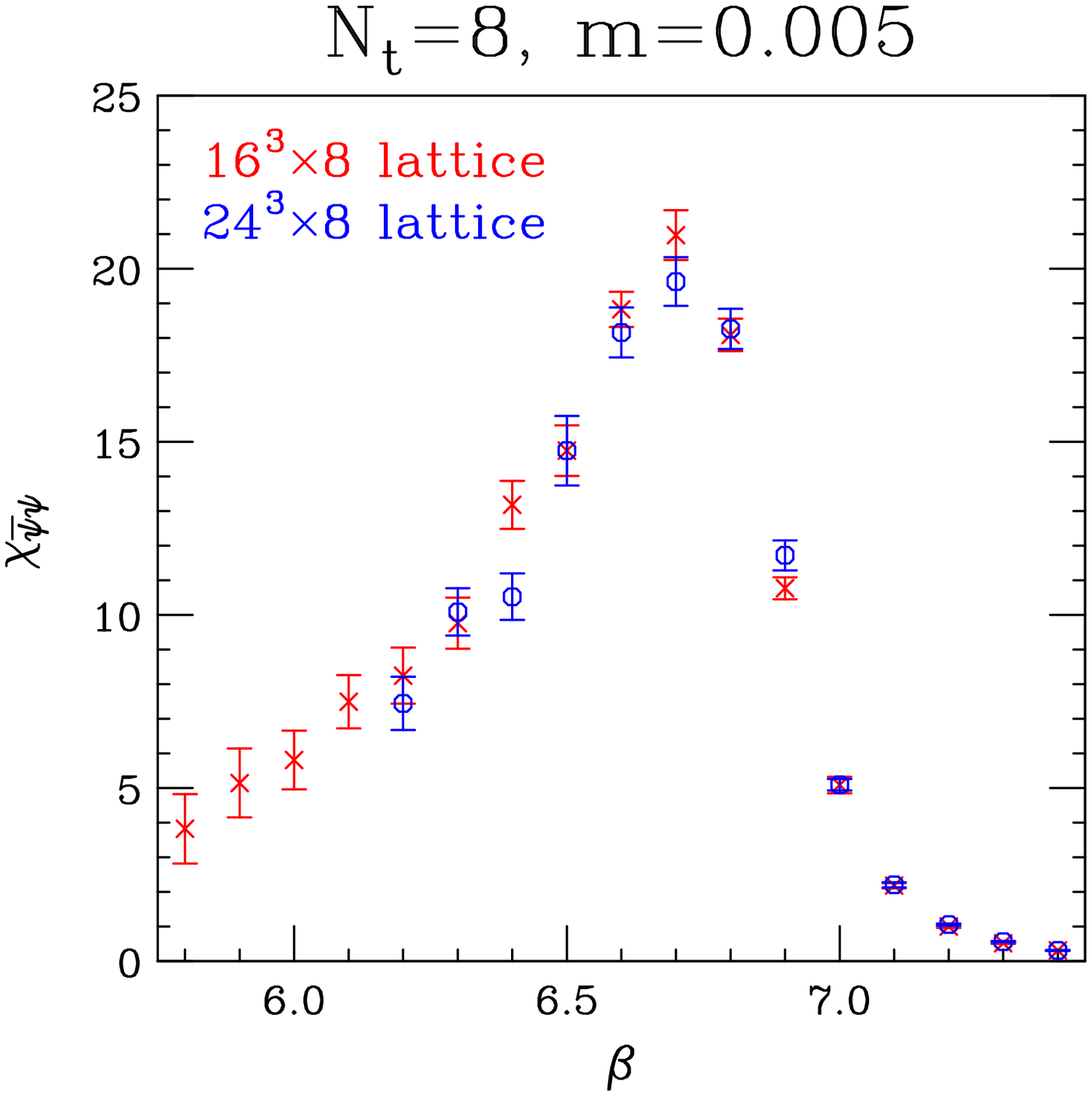}
\caption{a) The chiral susceptibilities as functions of $\beta$ for each of
the 3 masses on a $16^3 \times 8$ lattice. \newline
b) The chiral susceptibilities at $m=0.005$ as functions of $\beta$ on a
$16^3 \times 8$ lattice and on a $24^3 \times 8$ lattice.}
\label{fig:chi}
\end{figure}

It is clear that the chiral condensate becomes small for large $\beta$,
and decreases with decreasing quark mass, which suggests that it will vanish
in the chiral limit, for $\beta$ large enough. However, extrapolating the chiral
condensate to zero quark mass to determine the chiral transition from these
quark masses where the $\beta$ dependence is so smooth would be exceedingly
difficult. We therefore estimate the position of the chiral transition from
determinations of the positions of the peaks in the chiral susceptibilities
for each mass. These are plotted in figure~\ref{fig:chi}a. For the lower two
masses, the peaks in the chiral susceptibilities are well defined. (This is
the best evidence we have that our quark masses are small enough to perform
the chiral extrapolation.) In addition, for the limited set of $\beta$ values
of our simulations, both the $m=0.01$ and the $m=0.005$ `data' peak at the
same $\beta$, namely $\beta=6.7$. We therefore estimate that the position of
the peak and thus the chiral phase transition at $m=0$ are at
$\beta_\chi=6.7(1)$. This means that $\beta_d$ and $\beta_\chi$ are well
separated as was observed for $N_t=4$ and $6$. At $m=0.005$, close to the
chiral transition, we have also performed simulations on larger ($24^3 \times
8$) lattices. The Wilson Lines and chiral condensates show little difference
between the two lattice sizes. The chiral susceptibilities plotted in
figure~\ref{fig:chi} indicate that finite size effects are indeed small. This
is more significant, since such fluctuation quantities are most sensitive to
finite volume effects.

We have also looked at the chiral susceptibilities for the states with real
negative or complex Wilson lines and find peaks at $\beta = 6.8(1)$ for
$m=0.02$, $\beta = 6.7(1)$ for $m=0.01$ and $\beta = 6.6(1)$ for $m=0.005$.
Since these values are close to the transitions from the state with a negative
Wilson Line to states with complex Wilson Lines, there is a possibility of
interference between these two transitions. For this reason we concentrate our
studies on the chiral transition measured in the positive Wilson Line state.

\begin{figure}[htb]
\epsfxsize=3.2in
\epsffile{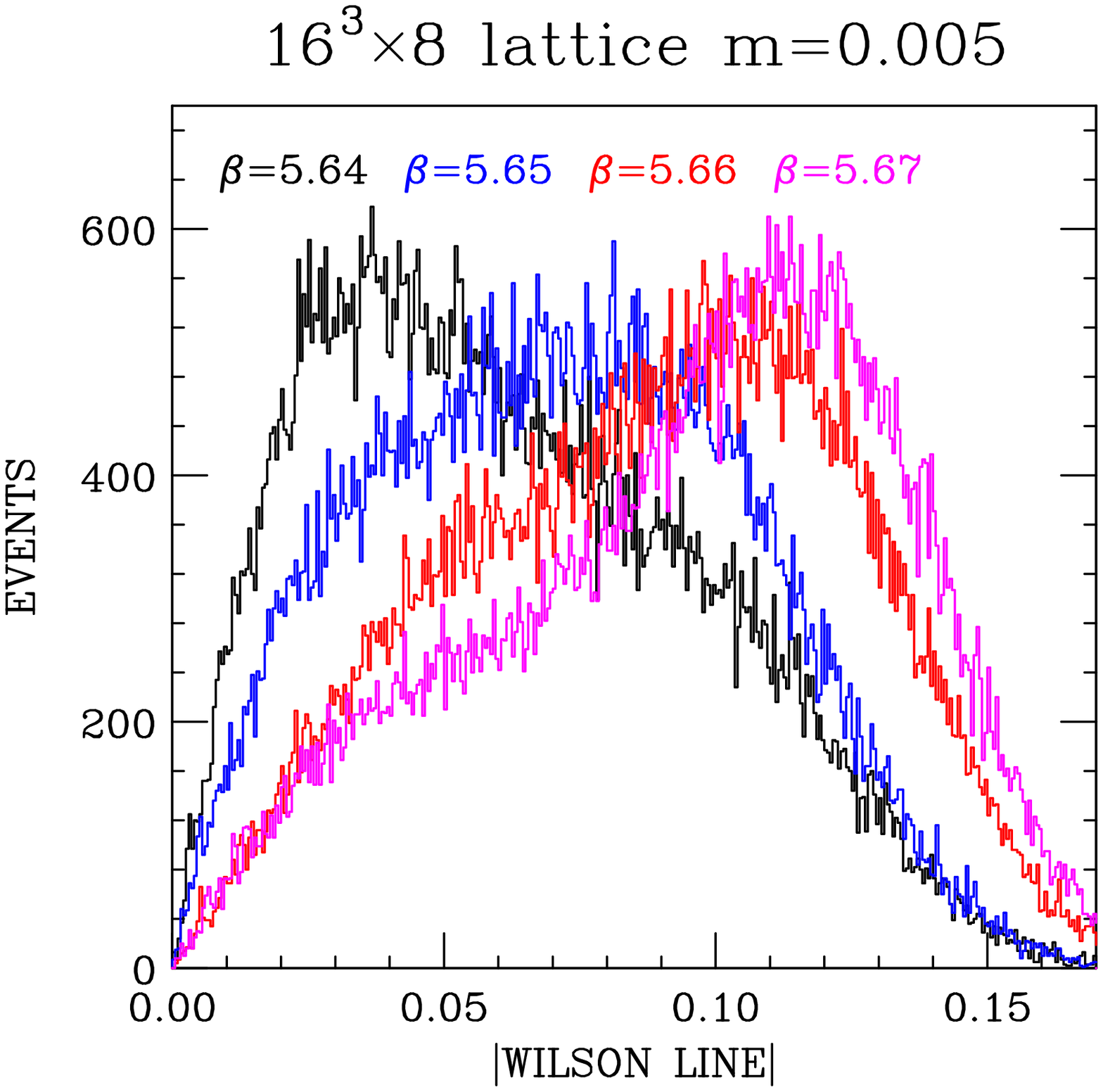}
\epsfxsize=3.2in
\epsffile{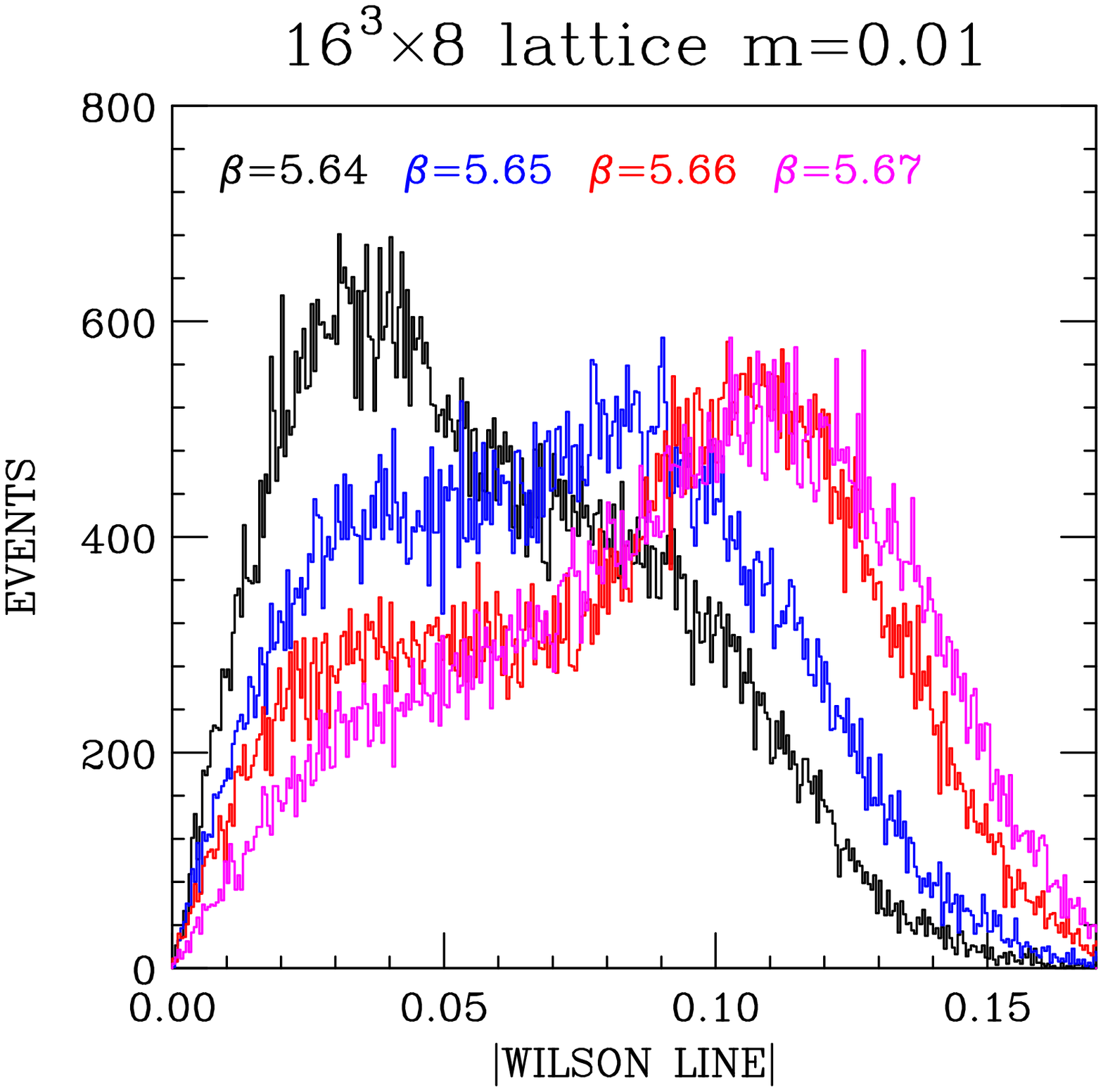}
\epsfxsize=3.2in                    
\epsffile{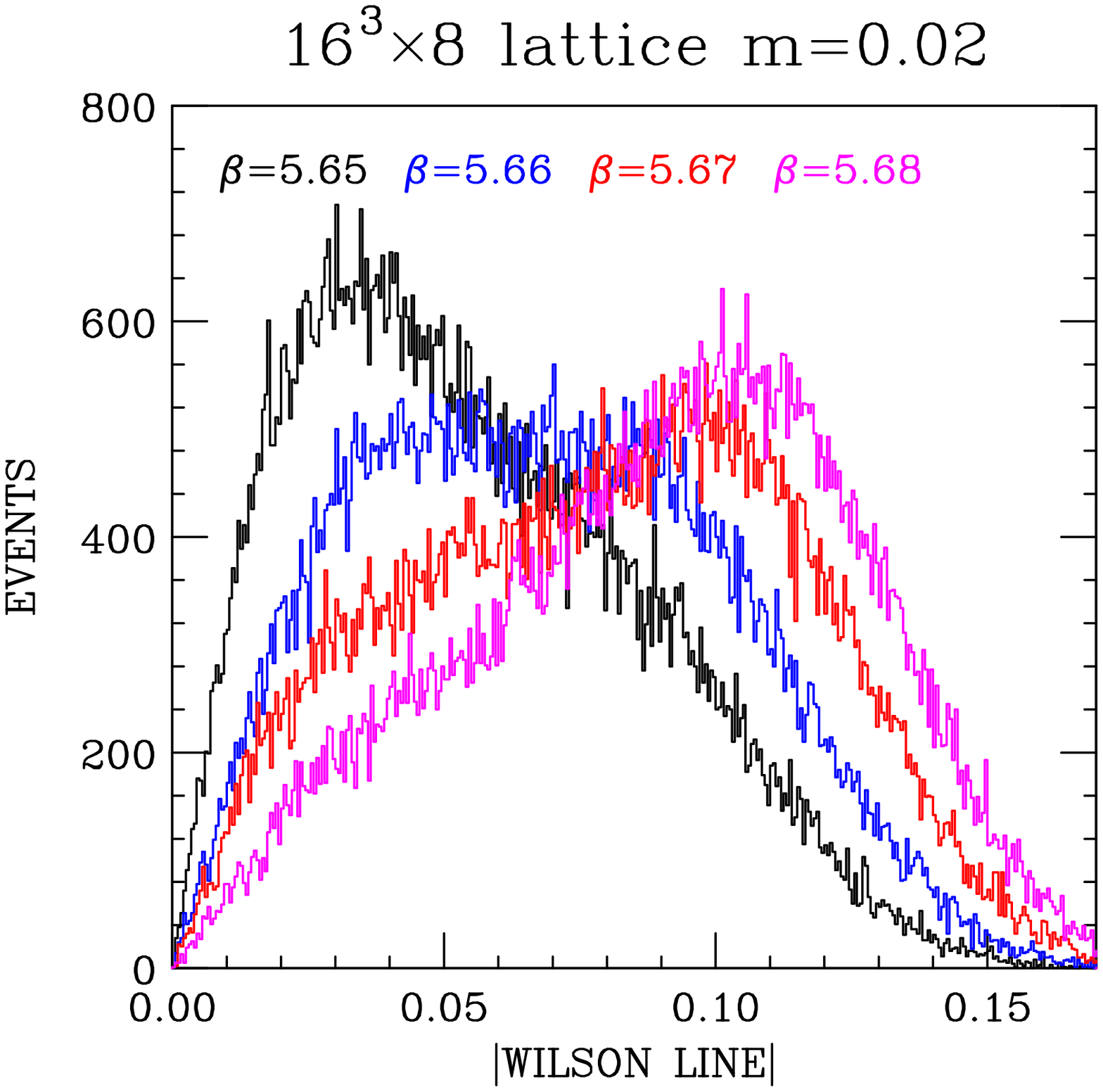}
\caption{Histograms of magnitudes of Wilson Lines: a) for $m=0.005$, b) for
$m=0.01$ and c) for $m=0.02$.}
\label{fig:hist}
\end{figure}

We now turn our attention to more precise estimates of $\beta_d$. For this
purpose, we histogram the magnitudes of the Wilson Lines in the neighbourhood
of the deconfinement transition. Such histograms are shown in 
figure~\ref{fig:hist} for each quark mass. We estimate that the transition
occurs at $\beta=\beta_d=5.66(1)$ for $m=0.02$, at $\beta_d=5.65(1)$ for 
$m=0.01$ and at $\beta_d=5.65(1)$ for $m=0.005$.

The positions of the deconfinement and chiral transitions, extrapolated to the 
chiral limit are given in table~\ref{tab:trans} for each of the 3 $N_t$ values
($N_t=4,6$ from \cite{Kogut:2010cz}, $N_t=8$ this work).
\begin{table}[htb]
%\parbox{0.25in}{$\:$}
%\parbox{2.5in}{
\centerline{
\begin{tabular}{|c|c|c|}
\hline
$N_t$          & $\beta_d$       & $\beta_\chi$             \\
\hline
4              &$\;$5.40(1)$\;$  &$\;$6.3(1)$\;$            \\
6              &$\;$5.54(1)$\;$  &$\;$6.6(1)$\;$            \\
8              &$\;$5.65(1)$\;$  &$\;$6.7(1)$\;$            \\
\hline
\end{tabular}
}
\caption{$N_f=2$ deconfinement and chiral transitions for $N_t=4,6,8$.}
\label{tab:trans}
\end{table}

\subsection{Interpretation}

We now compare the changes in $\beta_d$ and $\beta_\chi$ with what would
be expected if the running of the lattice coupling constant is given by
the 2-loop (perturbative) $\beta$ function -- equation~\ref{eqn:deltabeta}.

For the deconfinement transition 
\begin{equation}
\Delta\beta_d(6,4)=\beta_d(N_t=6)-\beta_d(N_t=4) \approx 0.14
\end{equation}
compared with the prediction of equation~\ref{eqn:deltabeta} which predicts
$\Delta\beta_d(6,4) \approx 0.12$, whereas
\begin{equation}                                                               
\Delta\beta_d(8,6)=\beta_d(N_t=8)-\beta_d(N_t=6) \approx 0.11 \: .
\end{equation}
compared with the 2-loop prediction $\Delta\beta_d(8,6) \approx 0.09$. If
the quarks were actively screening colour at these couplings 
($5.40 \lesssim \beta \lesssim 5.65$), it would not be unreasonable to
assume that these deconfinement couplings were weak enough to be governed
by the perturbative $\beta$ function. The fact that the measured $\Delta\beta$s
are within $\approx$~20\% of those predicted by 2-loop perturbation theory
would tend to support this interpretation. However, examining the running of
the coupling constant at the chiral transition, will lead us to a different
conclusion.

For the chiral transition, we find
\begin{equation}                                                       
\Delta\beta_\chi(6,4)=\beta_\chi(N_t=6)-\beta_\chi(N_t=4) \approx 0.3
\end{equation} 
while 
\begin{equation}
\Delta\beta_\chi(8,6)=\beta_\chi(N_t=8)-\beta_\chi(N_t=6) \approx 0.1 \: .
\end{equation}  
Using equation~\ref{eqn:deltabeta} to estimate $\Delta\beta_\chi$, yields
\begin{equation}
\Delta\beta_\chi(6,4) \approx 0.122
\end{equation}
and
\begin{equation}                                                        
\Delta\beta_\chi(8,6) \approx 0.087   .                                       
\end{equation}  
While $\Delta\beta_\chi(8,6)$ appears consistent with our measurements,
$\Delta\beta_\chi(6,4)$ does not. What this suggests is that for $N_t$ in the
range 6--8, $\beta_\chi$ is in the weakly-coupled domain where scaling is
controlled by asymptotic freedom, while $N_t$ in the range 4--6 is in the
strongly coupled domain. 

In the strongly coupled domain, the fermions have formed a chiral condensate,
which effectively stops them from contributing significantly to the running of
the coupling constant. Hence we expect that the running of the coupling in
this region will be that of the quenched theory, i.e. that for
equation~\ref{eqn:deltabeta} with $N_f=0$. This yields 
\begin{equation}                                                               
\Delta\beta_\chi(6,4) \approx 0.357 ,
\end{equation}                                                                  
which is consistent with what we observe. (It also gives
\begin{equation}                                                             
\Delta\beta_\chi(8,6) \approx 0.253 ,                                   
\end{equation}
which is larger than what we observe.)
Thus we conclude that the chiral transition emerges from the strongly coupled
domain, where the quarks play little part in the coupling constant evolution,
into the weak coupling regime, where the running of the coupling is determined
by asymptotic freedom, around $\beta_\chi(N_t=6)$.

\begin{figure}[htb]
\epsfxsize=5.0in
\epsffile{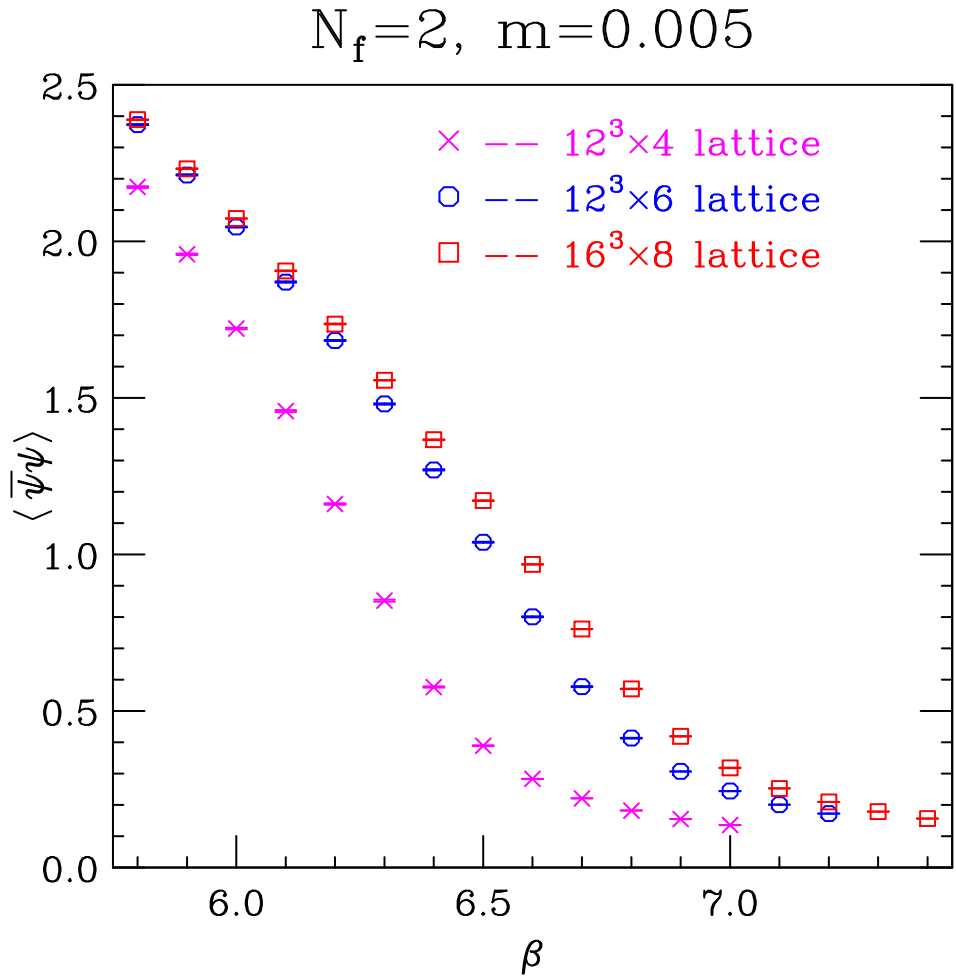}
\caption{Chiral condensates for $m=0.005$ for $N_t=4$, $6$ and $8$, in the
high $\beta$ (weak coupling) regime.}
\label{fig:psiNt}
\end{figure}

One might argue that both $\Delta\beta_\chi$s are consistent with either 
$N_f=0$ or $N_f=2$ scaling, because of the relatively large error-bars in
table~\ref{tab:trans}. However, comparing the graphs of the chiral condensates
for fixed masses for each $N_t$ -- see figure~\ref{fig:psiNt} -- we see that
$\Delta\beta_\chi(6,4)$ really does appear to be much larger than
$\Delta\beta_\chi(8,6)$, and that the estimates of equations 11 and 12 are
more accurate than the errors in the individual $\beta_\chi$s would suggest.

This interpretation of the running of the lattice coupling constant at the
chiral transition indicates that the region $\beta \lesssim 6.6$ is one of
strong coupling, governed
by quenched dynamics. In particular, the change in $\beta_d$ as $N_t$ is varied
from $4$
to $6$ to $8$ should be governed by quenched dynamics. However, it has been
determined that the evolution of the deconfinement coupling in quenched QCD,
is only well described by 2-loop perturbation theory for $\beta \gtrsim 6.1$
\cite{Gottlieb:1985ug,Christ:1985wx}. Hence the changes in $\beta_d$ that we
observe are not expected to be described by quenched perturbation theory. We
note, however, that $\beta_d(N_t=6)=5.54(1)$ is close to $\beta_d(N_t=3)
\approx 5.55$ found for the quenched theory \cite{Celik:1983gu}, while
$\beta_d(N_t=8)=5.65(1)$, compared to $\beta_d(N_t=4) \approx 5.69$ for the
quenched theory \cite{Celik:1983gu,Kennedy:1984dk,Brown:1988qe}. Since the
ratios of lattice spacings in these 2 cases are identical, such comparison is
justified. Taking into account the fact that small $N_t$ effects are expected
to make the $N_t=3$ quenched $\beta_d$ anomalously small, the comparison is
remarkably good. Unfortunately we cannot expect similar comparisons between
$\beta_d(N_t=4)$ and $\beta_d(N_t=2)$ for the quenched theory, to work, since
it is well known that $\beta_d(N_t=2) \approx 5.1$ for quenched lattice QCD, is
anomalously low \cite{Celik:1983gu}. Still it is reasonable that a
strong-coupling quenched interpretation of the running of $\beta_d$ is correct
for $N_t=4,6,8$. The fact that the changes in $\beta_d$ between $N_t=4,6$ and
$8$ are appreciably less than predicted by quenched perturbation theory is a
well-known feature of the strong-coupling domain of quenched lattice QCD
\cite{Gottlieb:1985ug,Christ:1985wx}.

\section{Discussion and Conclusions.}

We are studying thermodynamics of QCD with 2 massless colour-sextet `quarks' in
an attempt to distinguish whether it is a conformal field theory or if it 
`walks', i.e. is a confined, chiral-symmetry broken theory, with a slowly
evolving coupling. Simulations are performed on lattices with spatial extent
$N_s a$ and temporal extent $N_t a$ ($a$ is the lattice spacing) with
$N_s >> N_t$. We use staggered quarks with several small masses to extrapolate
to the massless quark limit. The temperature $T=1/N_t a$. Hence for fixed $T$,
chosen to be either the deconfinement temperature $T_d$ or the chiral-symmetry
restoration temperature $T_\chi$, we can vary $a$ by varying $N_t$, thus
studying the running of the coupling $g(a)$ as $a \rightarrow 0$ by simulating
at a series of $N_t$s increasing towards infinity. Our earlier simulations
were performed using $N_t=4,6$. Those we report here use $N_t=8$.

At $N_t=8$, as at the smaller $N_t$s, the chiral transition occurs at a much
weaker coupling than the deconfinement transition (see table~\ref{tab:trans}).
(This contrasts with the case with quarks in the fundamental representation of
colour, where these transitions appear coincident.) Between $N_t=4$ and $6$
both transitions move to appreciably smaller couplings. While this trend
continues between $N_t=6$ and $8$, the change in couplings at the chiral
transition, is much smaller than that between $N_t=4$ and $6$. A possible
explanation is that for couplings between those at the $N_t=4$ and the $N_t=6$
chiral transitions, the system is in the strong-coupling regime, where the
quarks are bound in a chiral condensate at distances $\lesssim a$, and thus do
not contribute significantly to the evolution of the coupling constant, which
thus evolves as in quenched ($N_f=0$) QCD. Between the couplings for the
$N_t=6$ and $N_t=8$ transitions, the system emerges into the weak-coupling
domain, where the fermions contribute and the coupling evolves according to
$N_f=2$ asymptotic freedom. Although we have given semi-quantitative evidence
for this interpretation, we cannot rule out the possibility that the coupling
at the chiral transition is approaching a fixed non-zero value. This would be
evidence for a bulk chiral transition, implying that the continuum theory has
unbroken chiral symmetry and is thus conformal.

In order to distinguish conformal from walking behaviour, we will need to
perform simulations at larger $N_t$ values. We have already begun simulations
on $N_t=12$ lattices. In addition, since the changes in $\beta_\chi$ between
$N_t=6$ and $N_t=8$ and those expected between $N_t=8$ and $N_t=12$ are of
order $0.1$ we will need more $\beta$ values in the neighbourhood of 
$\beta_\chi$ to determine this value more accurately. We are currently 
performing such simulations and additional simulations with a smaller quark
mass ($m=0.0025$) at $N_t=8$, to aid with the chiral extrapolation. With these
new simulations, we are concentrating on the chiral transition, since it
would require $N_t$ values much greater than what is currently feasible to
have $\beta_d$ in the weak-coupling ($\beta \gtrsim 6.6$) regime.

Our runs at $N_t=8$ show a phase structure similar to what was observed at
$N_t=6$. Above the deconfinement transition, the Wilson Line exhibits a definite
3-state structure. In addition to the state with a positive Wilson Line, which
is all that is observed for fundamental quarks, there are states characterized
by Wilson Lines oriented (at least approximately) in the directions of the
2 complex cube roots of unity. The existence of this 3-state signal is probably
because chiral symmetry is still broken in this regime, effectively removing
the fermions from the dynamics, so that it behaves as a quenched theory. This
suggests that the 3-state signal is the vestage of the spontaneously broken
$Z_3$ symmetry of the deconfined pure gauge theory. The fact that this $Z_3$
symmetry is explicitly broken manifests itself in the fact that the magnitude
of the Wilson Line in the complex Wilson Line states is smaller than that in
the positive Wilson Line state. Within the limitations of our simulations, all
3 states appear to be stable. At some $\beta$ value close to the chiral
transition, the complex Wilson Line states disorder to a state with a negative
Wilson Line. Above this transition the magnitude of the Wilson Line in the
negative Wilson Line state is around $1/3$ of that for the positive Wilson
line state. This leads us to speculate that the transition indicates a
breaking of colour $SU(3)$ to colour $SU(2) \times U(1)_Y$. Arguments for the
existence of states with Wilson Lines having phases $\pm 2\pi/3$ and $\pi$ in
addition to that with phase $0$ have been given by Machtey and Svetitsky, who
showed evidence for them in their simulations with Wilson quarks.

%%% REWRITE
We also plan simulations to measure the zero-temperature properties of this
theory. In the weak-coupling regime $\beta > \beta_\chi$, we will check whether
the theory is a conformal field theory or if it is in the quark-gluon plasma
phase of a QCD-like gauge theory. If the theory is a conformal field theory
(for massless fermions) all `hadron' masses will vanish with the same anomalous
dimension, and the chiral condensate will also vanish with an anomalous
dimension, in the chiral limit. Because such anomalous dimensions are determined
by the infrared attractive fixed-point, they should be independent of $\beta$.

If we do not find evidence of conformal behaviour, we will check for QCD-like
behaviour in the chirally-broken phase, and for evidence that this phase has a
continuum limit controlled by asymptotic freedom. This will require very large
lattices, since we need to choose $\beta$ values large enough for asymptotic
freedom to control the renormalization group scaling of observables, while
keeping $\beta < \beta_d (<\beta_\chi)$. Here we will need to measure the
masses of the `hadrons' to determine if our quark masses are small enough and
our lattices large enough to observe that the `pion' masses vanish in the
chiral limit proportional to $\sqrt{m}$, while the other `hadrons' remain
massive. We will also need to check for evidence that the chiral condensate
remains finite in the chiral limit. In addition we will measure $f_\pi$ and
study propagators of vector and axial vector mesons which contribute to the
$S$ parameter, as is being done by the Lattice Strong Dynamics Collaboration,
for fundamental quarks \cite{Appelquist:2010xv}. Simulations will be
performed at several $\beta$ values to determine the running of the bare
coupling and of some appropriately-defined renormalized coupling. This is
necessary to check that the theory has the correct ultraviolet completion.

Zero temperature simulations with sextet quarks are already being performed
using improved staggered quarks by the Lattice Higgs Collaboration, who
presented preliminary results at Lattice 2010 \cite{Fodor:2011tw}.

\section*{Acknowledgements}

DKS is supported in part by the U.S. Department of Energy, Division of High
Energy Physics, Contract DE-AC02-06CH11357. JBK is supported in part
by NSF grant NSF PHY03-04252. These simulations were performed on the Linux
Cluster, Fusion, at the LCRC at Argonne National Laboratory, and the Linux
Cluster Carver/Magellan at NERSC under an ERCAP allocation.

DKS thanks J.~Kuti, D.~Nogradi, F.~Sannino, J.~Giedt and B.~Svetitsky for
informative discussions. We thank D.~Nogradi of the Lattice Higgs
Collaboration for using their code to perform an independent check of some of
our small-lattice results.

\appendix

\section{Run details}

Table~\ref{tab:runs16} gives the length of our $16^3 \times 8$ runs in
length-1 trajectories, for each $\beta$ and mass. Where 2 numbers are given,
the first is for a series of runs which started from an ordered configuration
at large $\beta$, while the second is from a start which gives negative Wilson
Loops at large $\beta$.

\begin{table}[htb]
\begin{tabular}{|d|c|c|c|}
\hline
\multicolumn{1}{|c|}
{$\beta$}     &     $m=0.005$      &      $m=0.01$      &     $m=0.02$       \\
\hline
5.5           &   10,000           &   10,000           &   10,000           \\
5.55          &   10,000           &   10,000           &   10,000           \\
5.6           &   50,000           &   50,000           &   50,000           \\
5.64          &   50,000 + 50,000  &   50,000 + 50,000  &    ---             \\
5.65          &   50,000 + 50,000  &   50,000 + 50,000  &   50,000 + 50,000  \\
5.66          &   50,000 + 50,000  &   50,000 + 50,000  &   50,000 + 50,000  \\
5.67          &   50,000 + 50,000  &   50,000 + 50,000  &   50,000 + 50,000  \\
5.68          &   50,000 + 50,000  &   50,000 + 50,000  &   50,000 + 50,000  \\
5.7           &   50,000 + 50,000  &   50,000 + 50,000  &   50,000 + 50,000  \\
5.8           &   10,000 + 10,000  &   10,000 + 10,000  &   10,000 + 10,000  \\
5.9           &   10,000 + 10,000  &   10,000 + 10,000  &   10,000 + 10,000  \\
6.0           &   10,000 + 10,000  &   10,000 + 10,000  &   10,000 + 10,000  \\
6.1           &   10,000 + 10,000  &   10,000 + 10,000  &   10,000 + 10,000  \\
6.2           &   10,000 + 10,000  &   10,000 + 10,000  &   10,000 + 10,000  \\
6.3           &   10,000 + 10,000  &   10,000 + 10,000  &   10,000 + 10,000  \\
6.4           &   10,000 + 10,000  &   10,000 + 10,000  &   10,000 + 10,000  \\
6.5           &   10,000 + 50,000  &   10,000 + 10,000  &   10,000 + 10,000  \\
6.6           &   20,000 + 50,000  &   10,000 + 50,000  &   10,000 + 10,000  \\
6.7           &   20,000 + 50,000  &   10,000 + 50,000  &   10,000 + 10,000  \\
6.8           &   20,000 + 50,000  &   10,000 + 50,000  &   10,000 + 50,000  \\
6.9           &   10,000 + 20,000  &   10,000 + 50,000  &   10,000 + 50,000  \\
7.0           &   10,000 + 10,000  &   10,000 + 50,000  &   10,000 + 50,000  \\
7.1           &   10,000 + 10,000  &   10,000 + 10,000  &   10,000 + 30,000  \\
7.2           &   10,000 + 10,000  &   10,000 + 10,000  &   10,000 + 10,000  \\
7.3           &   10,000 + 10,000  &   10,000 + 10,000  &   10,000 + 10,000  \\
7.4           &   10,000 + 10,000  &   10,000 + 10,000  &   10,000 + 10,000  \\
\hline 
\end{tabular}
\caption{Numbers of trajectories for each $\beta$ and $m$ for runs on 
$16^3 \times 8$ lattices. The first number is for simulations starting with
positive Wilson Lines; the second is for simulations starting with negative or
complex Wilson Lines.}
\label{tab:runs16}
\end{table}

\end{document}